\documentclass[useAMS,usenatbib]{mnras}
\usepackage{graphicx} 
\usepackage{subcaption}
\usepackage{amsmath} 
\usepackage{xcolor}
\usepackage{float}
\usepackage{caption}
\usepackage{placeins}
\usepackage{lipsum}
\usepackage{multicol}
\usepackage{soul}
\usepackage{amssymb}
\usepackage{ragged2e}
\usepackage{threeparttable}

\def\refjnl#1{{\rm#1}}
\def\aj{\refjnl{AJ}}                   
\def\apj{\refjnl{ApJ}}                 
\def\apjl{\refjnl{ApJ}}                
\def\apjs{\refjnl{ApJS}}               
\def\ao{\refjnl{Appl.~Opt.}}           
\def\aap{\refjnl{A\&A}}                
\def\jcap{\refjnl{J. Cosmology Astropart. Phys.}} 
\def\mnras{\refjnl{MNRAS}}             
\def\prd{\refjnl{Phys.~Rev.~D}}        
\def\prl{\refjnl{Phys.~Rev.~Lett.}}    
\def\nat{\refjnl{Nature}}              
\def\physrep{\refjnl{Phys.~Rep.}}   

\def\be{\begin{equation}} 
\def\ee{\end{equation}}

\def\msun{{\Msun}}

\def\gsim{\lower.5ex\hbox{\gtsima}} 
\def\lsim{\lower.5ex\hbox{\ltsima}} \def\gtsima{$\; \buildrel > \over 
\sim \;$} \def\ltsima{$\; \buildrel < \over \sim \;$} \def\prosima{$\; 
\buildrel \propto \over \sim \;$} \def\gsim{\lower.5ex\hbox{\gtsima}} 
\def\lsim{\lower.5ex\hbox{\ltsima}} 
\def\simgt{\lower.5ex\hbox{\gtsima}} 
\def\simlt{\lower.5ex\hbox{\ltsima}} 
\def\simpr{\lower.5ex\hbox{\prosima}} \def\la{\lsim} \def\ga{\gsim} 
  
 \def\gtsima{$\; \buildrel > \over \sim \;$} 
\def\ltsima{$\; \buildrel < \over \sim \;$} 
\def\gsim{\lower.5ex\hbox{\gtsima}} 
\def\lsim{\lower.5ex\hbox{\ltsima}} 
\def\simgt{\lower.5ex\hbox{\gtsima}} 
\def\simlt{\lower.5ex\hbox{\ltsima}} 
\def\simpr{\lower.5ex\hbox{\prosima}} 
\def\la{\lsim} 
\def\ga{\gsim}

\def\msun{\,{\rm \Msun}}

\def\E3{{\cal E}_{\rm g}^{III}}

\def\msun{\rm M_\odot}

\def\cmpct{\rm cMpc^{-3}}
\def\M*{M_*}

\def\Z*{Z_*}
\def\L*{L_*}

\def\mx{\,m_x}

\title[WDM constraints from the JWST]{Warm dark matter constraints from the JWST} 
\author[Dayal \& Giri]{Pratika Dayal$^{1}$\thanks{p.dayal@rug.nl} \& Sambit K. Giri$^2$ \\ 
$^{{1}}$Kapteyn Astronomical Institute, University of Groningen, PO Box 800, 9700 AV Groningen, The Netherlands  \\
$^2$Nordita, KTH Royal Institute of Technology and Stockholm University, Hannes Alf\'vens v\"ag 12, SE-106 91 Stockholm, Sweden \\}

\begin{document} 
 
\date{} 

\maketitle

\begin{abstract}
Warm Dark Matter (WDM) particles with masses ($\sim$ kilo electronvolt) offer an attractive solution to the small-scale issues faced by the Cold Dark Matter (CDM) paradigm. The delay of structure formation in WDM models and the associated dearth of low-mass systems at high-redshifts makes this an ideal time to revisit WDM constraints in light of the unprecedented data-sets from the James Webb Space Telescope (JWST). Developing a phenomenological model based on the halo mass functions in CDM and WDM models, we calculate high-redshift ($z \gsim 6$) the stellar mass functions (SMF) and the associated stellar mass density (SMD) and the maximum stellar mass allowed in a given volume. We find that: (i) WDM as light as 1.5 keV is already disfavoured by the low-mass end of the SMF (stellar mass $M_* \sim 10^7\msun$) although caution must be exerted given the impact of lensing uncertainties; (ii) 1.5 keV WDM models predict SMD values that show a steep decrease from $10^{8.8}$ to $10^{2} \msun \cmpct$ from $z \sim 4$ to 17 for $M_* \gsim 10^8 \msun$; (iii) the 1.5 keV WDM model predicts a sharp and earlier cut-off in the maximum stellar masses for a given number density (or volume) as compared to CDM or heavier WDM models. For example, with a number density of $10^{-3} \rm {cMpc^{-3}}$, 1.5 (3) KeV WDM models do not predict bound objects at $z \gsim 12$ (18). Forthcoming JWST observations of multiple blank fields can therefore be used as a strong probe of WDM at an epoch inaccessible by other means. 

\end{abstract}


\section{Introduction}
The nature of dark matter (DM) remains a key outstanding question in the field of physical cosmology. Small-scale issues of the standard cold DM (CDM) paradigm \citep[for a review see e.g.][]{weinberg2013} have motivated an entire zoo of ``DM candidates beyond CDM" including Warm DM \citep[WDM; e.g. ][]{blumenthal1982, bode2001}, fuzzy DM (FDM) consisting of ultra-light ($\sim 10^{-22}$ electronvolt) boson or scalar particles \citep{hu2000, marsh2014}, interacting DM \citep{spergel2000} and decaying DM \citep{wang2014}. A key property of these ``beyond-CDM" models is their ability to smear out small-scale power, thereby suppressing the formation of low-mass structures. Focusing on WDM, the Lyman Alpha (Ly$\alpha$) forest power spectrum measured from high-resolution quasar spectra at redshifts $z \sim 2-5$ yield a WDM particle mass $\mx \gsim 1.9-3.9$ keV \citep[e.g.][]{viel2013, irsic2017,garzilli2021, villasenor2022} depending on the exact assumptions made regarding the intergalactic medium (IGM) temperature-density relation. Further, the image positions and flux ratios of gravitationally lensed quasars have been used to infer $\mx \gsim 5.2-5.58$ keV \citep{hsueh2020, gilman2020} while combinations of strong lensing and Milky Way satellite populations have been used to infer the most stringent constraints with $\mx \gsim 6.04-9.7$ keV \citep{enzi2021, nadler2021}. 
 
 \begin{figure*}
\center{\includegraphics[scale=1.01]{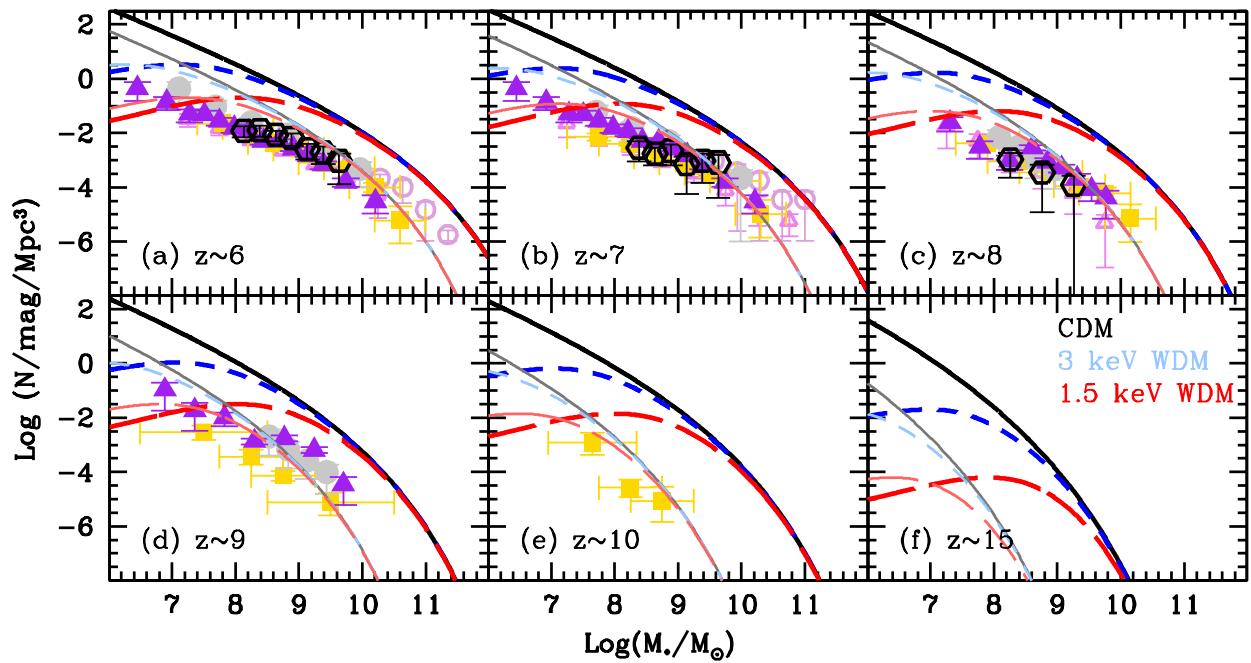}} 
\caption{The stellar mass function (SMF) at $z \sim 6-15$, as marked in panels (a)-(f). In each panel, the solid (black), short-dashed (blue) and long-dashed (red) lines show the SMFs for CDM, 3 keV and 1.5 keV WDM models, as marked in panel (f). Thin and thick lines show the results assuming $\epsilon_* = fn(z)$ and $\epsilon_*=1$ (models A and B; sec. 2.1); the latter model shows a physical upper limit to the SMF. Points show the observed data-sets from \citet[][open circles at $z \sim 6-7$]{duncan2014}, \citet[][open triangles at $z \sim 6-7$]{song2016}, \citet[][filled circles at $z \sim 6-9$]{bhatawdekar2019}, \citet[][filled triangles at $z \sim 6-9$]{kikuchihara2020}, \citet[][filled squares at $z \sim 6-10$]{stefanon2021} and recent JWST estimates from \citet[][empty hexagons]{navarro-carrera2023}. All these datasets have been renormalised to a Salpeter IMF \citep{stefanon2021}. The last (f) panel shows the upper limit to the predicted SMFs at $z \sim 15$.  }
\label{fig_smf}
\end{figure*}

Since structure formation proceeds hierarchically and WDM smears-out small-scale power, the effects of WDM are expected to be manifested most strongly through a decrease in the number density of low-mass haloes and an associated delay in the assembly of more-massive systems. A growing body of work is focusing on extending constraints on $\mx$ into the first billion years at $z>7$: analytic models, based on the halo mass function, have been compared to high-redshift observations to obtain constraints of $\mx > 1.3-2.9$ keV \citep{schultz2014high,menci2016stringent,corasaniti2017constraints,rudakovskyi2021constraints} with others making predictions of the ultra-violet luminosity function (UV LF) and redshift evolution of the stellar mass density (SMD) that can be tested with the James Webb Space Telescope \citep[JWST; e.g.][]{dayal2015,stoychev2019clues,lapi2022astroparticle,lovell2018, kurmus2022, maio2023}. Further, the reported detection of a global 21cm signal by the EDGES (Experiment to Detect the Global Epoch of Reionization Signature) collaboration from the epoch of reionization \citep[EoR; although see][]{singh2022}  
has been used to constrain $\mx \gsim 3-6.1$ keV \citep[e.g.][]{schneider2018, chatterjee2019}. These are now being supplemented by a new generation of small-scale ($\sim 20^3$ Mpc$^3$) hydrodynamic \citep[e.g.][]{maio2015first,villanueva2018warm} and zoom-in simulations \citep[e.g.][]{stoychev2019clues}. A key caveat, however, is that uncertainties in baryonic physics remain degenerate with the underlying DM model used \citep[for a discussion see e.g.][]{dayal2018,villanueva2018warm,Giri2021baryonicfeedback}. 

Over the past few months, the JWST has provided unprecedented views of galaxy formation in the first billion years, yielding a number of galaxy candidates between $z \sim 9-16.5$ given its exquisite sensitivity \citep{bradley2022, donnan2022, atek2022, naidu2022a, adams2022, austin2023}. Indeed, the evolving (rest-frame 1500\AA) UV LF has now been mapped out between $z \sim 5-16.5$ \citep[e.g.][]{harikane2021b, bouwens2021,naidu2022a,harikane2022, bouwens2022jwst} although caution must be exerted when using the LF at $z \gsim 12$ where the redshift and nature of the sources remain debated \citep[e.g.][]{adams2022, naidu2022, arrabal-haro2023}. In terms of stellar mass, the JWST has yielded galaxy candidates with masses between $10^{8} -10^{11}\msun$ at $z \sim 7-10$ \citep{labbe2023, navarro-carrera2023}.

In this work, our aim is to use the latest data sets from the JWST to revisit constraints on the WDM particle mass in the first billion years using a simple phenomenological model that links galaxy stellar masses to their host DM haloes. We note that the analysis carried out here can be applied to any DM model for which halo mass functions (HMFs) can be calculated within the first billion years. 

We present our methodology for calculating the HMFs for different cosmologies in Sec. \ref{model} before presenting our phenomenological model for the star formation efficiency in Sec. \ref{stel}. We compare our theoretical stellar mass functions (SMFs) and the associated stellar mass density (SMD) to observations in Sec. \ref{smf} before discussing the maximum stellar mass allowed in any DM model for a given number density (or volume) in Sec. \ref{maxms} before concluding in Sec. \ref{conc}.
Finally, throughout this work, we assume cosmological parameters in accord with \citet{planck2018} such that $\Omega_m=0.315$, $\Omega_b=0.049$, $\sigma_8=0.813$, $n_s=0.963$ and $h_0=0.673$. We quote all length scales in comoving Mpc (cMpc).

\begin{figure*}
\center{\includegraphics[scale=0.9]{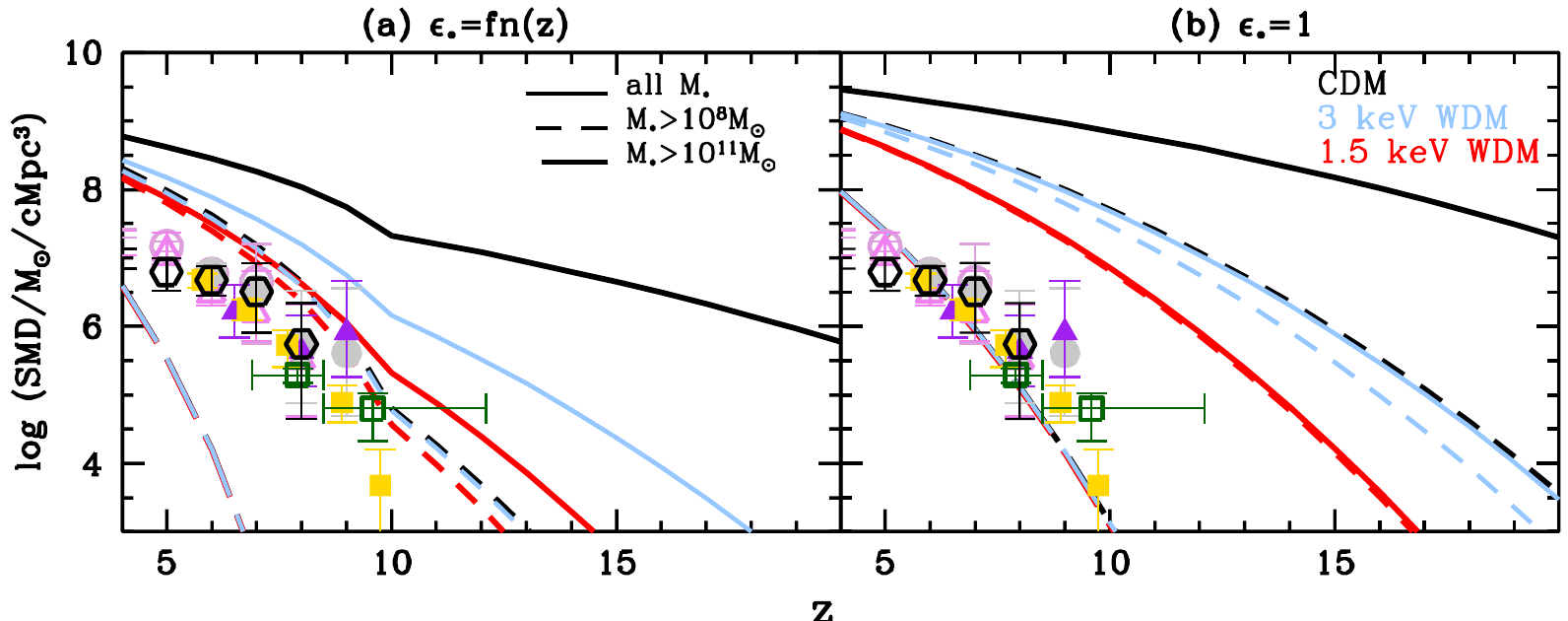}} 
\caption{The redshift evolution of the stellar mass density (SMD) from $z \sim 4-20$ for both star formation efficiency models considered in this work (see Sec. 2.1): panel (a) shows $\epsilon_* =fn(z)$ while panel (b) shows results for $\epsilon_*=1$, the physical upper limit to the SMD. As marked, black, blue and red lines show the results for CDM, 3 keV and 1.5 keV WDM, respectively. Lines show results for all galaxies (solid lines) and integrating down to stellar mass limits of: $10^8 \msun$ (short-dashed lines) and $10^{11} \msun$ (long-dashed lines). For reference, points show the observed SMD inferred by integrating down to $10^8 \msun$ by \citet[][open circles]{duncan2014}, \citet[][open triangles]{song2016}, \citet[][filled circle]{bhatawdekar2019}, \citet[][filled triangles]{kikuchihara2020} and \citet[][filled squares]{stefanon2021}; finally, the empty squares show the latest JWST GLASS Early Release Science program results from \citet{santini2023} and from \citet[][empty hexagons]{navarro-carrera2023}.  }
\label{fig_smd}
\end{figure*}
\section{The theoretical model}
\label{model}
In this work, we use a phenomenological model based on the evolving HMF. Our aim is to explore constraints on the WDM particle mass using the (maximal) stellar masses allowed in different DM cosmologies. We start by constructing the HMF with the extended Press-Schechter (EPS) approach \citep[e.g.][]{giri2022imprints}. In brief, the halo number density $n$ per unit halo mass ($M_h$) is defined as,
\begin{eqnarray}
\frac{\mathrm{d} n}{\mathrm{d} \mathrm{log} M_h} =-\frac{\bar{\rho}}{M_h} f(\nu) \frac{\mathrm{d} \mathrm{log}\sigma}{\mathrm{d} \mathrm{log} M_h} \ ,
\end{eqnarray}
where $\nu \equiv \delta_c(z)/\sigma(R)$, $R$ is the halo radius and $\bar{\rho}$ is the mean background density. Further, $\delta_c(z) \approx 1.686/D(z)$, where $D(z)$ is linear growth factor, and $M_h=(4\pi/3)\bar{\rho}(cR)^3$, where $c=3.3$ \citep{parimbelli2021mixed}. 
We use the Sheth-Tormen form \citep{sheth1999large} for the first-crossing distribution $f(\nu)$, which can be written as,
\begin{eqnarray}
f(\nu) = A\sqrt{\frac{2\nu^2}{\pi}}(1+\nu^{-2p})e^{-\nu^2/2} \ ,
\end{eqnarray}
with $A=0.3222$ and $p=0.3$. The variance is given by
\begin{eqnarray}
\sigma^2(R,z) = \int \frac{k^2}{2\pi^2} P_\mathrm{L}(k,z)W^2_\mathrm{F}(k,R)~\mathrm{d}k \ ,
\end{eqnarray}
where $P_\mathrm{L}$ and $W_\mathrm{F}(k, R)$ are the linear matter power spectrum and the \textit{smooth-k} window function respectively. 
This window function is defined as $[1+(kR)^\beta]^{-1}$, where $\beta$ is set to 4.8 \citep{leo2018new,parimbelli2021mixed}. \citet{leo2018new} showed that the \textit{smooth-k} filter effectively addresses the limitations of both \textit{top-hat} and \textit{sharp-k} window functions in accurately reproducing the HMF derived from WDM $N$-body simulations, particularly at low masses.
We use the publicly available code, \textsc{class} \citep{lesgourgues2011classIV}, to model $P_\mathrm{L}$ for the cosmologies (CDM 
and $\mx \sim 1.5$ and 3 keV WDM) studied in this work. 

The suppression scale in the WDM power spectrum ($P_\mathrm{L}$) can be quantified using the half-mode length scale \citep[e.g.][]{schneider2012non},
\begin{equation}
    \lambda_\mathrm{hm} \approx 1.015 \bigg(\frac{\mx}{\rm keV}\bigg)^{-1.11} \bigg(\frac{\Omega_x}{0.25}\bigg)^{0.11} \bigg(\frac{h}{0.7}\bigg)^{1.22}  \mathrm{cMpc}.
\end{equation}
Here, $\Omega_x$ is the WDM background overdensity and $\lambda_\mathrm{hm}$ is the scale below which the WDM power spectrum is suppressed 4 times below CDM. 
This length scale corresponds to the half-mode mass scale given as $M_\mathrm{hm}=\frac{4\pi}{3}\rho_m\left(\frac{\lambda_\mathrm{hm}}{2}\right)$, where $\rho_m$ is the mean matter density.
For values of $\mx \sim 1.5$ (3) keV, $ \lambda_\mathrm{hm}$ is found to be $\sim$0.6 (0.3) cMpc, respectively, at $z=0$. In WDM cosmologies with $\mx \sim 1.5$ (3) keV, the number of haloes is suppressed below the half-mode mass $M_\mathrm{hm} \sim 10^{9.7}$ ($10^{8.7}$) $\msun$ at $z=6-15$.

We refer interested readers to \citet{schneider2012non} for a detailed comparison of the analytical formalism used here to $N$-body simulations.

\subsection{The stellar masses of early galaxies in different cosmologies}
\label{stel}
These HMFs above are used to obtain SMFs and the associated SMD at $z \sim 4-20$ as now detailed. We start by assuming 
each halo to contain gas mass that is linked to the halo mass through the cosmological ratio such that $M_g = (\Omega_b/\Omega_m)M_h$. We then study two different star formation efficiency ($\epsilon_*$) models to calculate the stellar mass as $M_* = \epsilon_* M_g$:
\begin{itemize}
\item Model A: In the first case, $\epsilon_*$ is chosen to match to the high-mass ($M_* \gsim 10^9 \msun$) end of the observed SMF at $z \sim 6-10$. This requires a star formation efficiency that decreases with increasing redshift such that $\epsilon_*= fn(z) = 0.15-0.03(z-6)$ at $z\sim 6-10$; we assume the star formation efficiency to saturate to $\epsilon_*= 0.03$ at $z>10$ given the paucity of observational data at these early epochs. 

\item Model B: In this {\it maximal} model, which presents a physical upper limit to the stellar mass contained in any halo, we assume a 100\% efficiency for the conversion of gas into stars i.e. $\epsilon_*=1$. This simple calculation yields an upper limit to the SMD and the maximum stellar mass possible in any given volume at each $z$. 
\end{itemize}

We use HMFs down to a minimum mass of $10^{6.5}\msun$ in all the cosmologies considered in this work. Since we do not include any suppression of the gas mass or star formation in our models, all of these haloes contribute when considering the ``total" SMD value at any redshift.

Observationally, the stellar mass is inferred by fitting to the observed photometry or spectroscopy, if available, and requires making a number of assumptions regarding the star formation history, the initial mass function (IMF), the dust attenuation, the impact of nebular emission lines and stellar binarity, to name a few. The assumed star formation history and the IMF can have significant consequences leading to stellar masses varying by as much as on order of magnitude \citep[e.g.][]{topping2022, wang2023} which can exceed the scatter from different spectral energy distribution (SED) fitting codes \citep{wang2023}. Theoretically, however, the stellar mass can directly be inferred from the gas mass assuming an efficiency of star formation. The associated mass-to-light ratio (M/L) depends on a number of parameters including the IMF, the age and metallicity of the stellar population and the dust content, to name a few. Indeed, for a given stellar mass, age and metallicity, the UV luminosity is higher by about an order of magnitude assuming a top-heavy IMF \citep[e.g.][with a slope of $-1$]{fardal2007} as compared to a Salpeter IMF \citep{salpeter1955} between $0.1-100 \msun$. On the other hand, the stellar mass calculated only depends on the assumptions of {\it (i)} a constant baryon-to-DM ratio; {\it (ii)} the redshift and mass dependence of $\epsilon_*$; and {\it (iii)} the assumed HMF evolution holding to redshifts as high as $z \sim 15-20$. This is why, in this work, we choose to compare stellar masses - we note that the observationally-inferred stellar masses used in this work have all been renormalised to a Salpeter IMF. We also use stellar masses since these offer a promising tool to be able to differentiate between CDM and non-CDM models \citep{dayal2015, kurmus2022, maio2023} and since a number of previous works \citep{dayal2015, lovell2018} have shown that luminosity-based indicators such as the UV LF are not a good probe of the underlying dark matter model. The higher mass-to-light ratios in non-CDM models partially compensate for the dearth of small-mass haloes, making the resulting UV LFs closer to CDM than expected from simple estimates of halo abundances.

\begin{figure*}
\center{\includegraphics[scale=0.75]{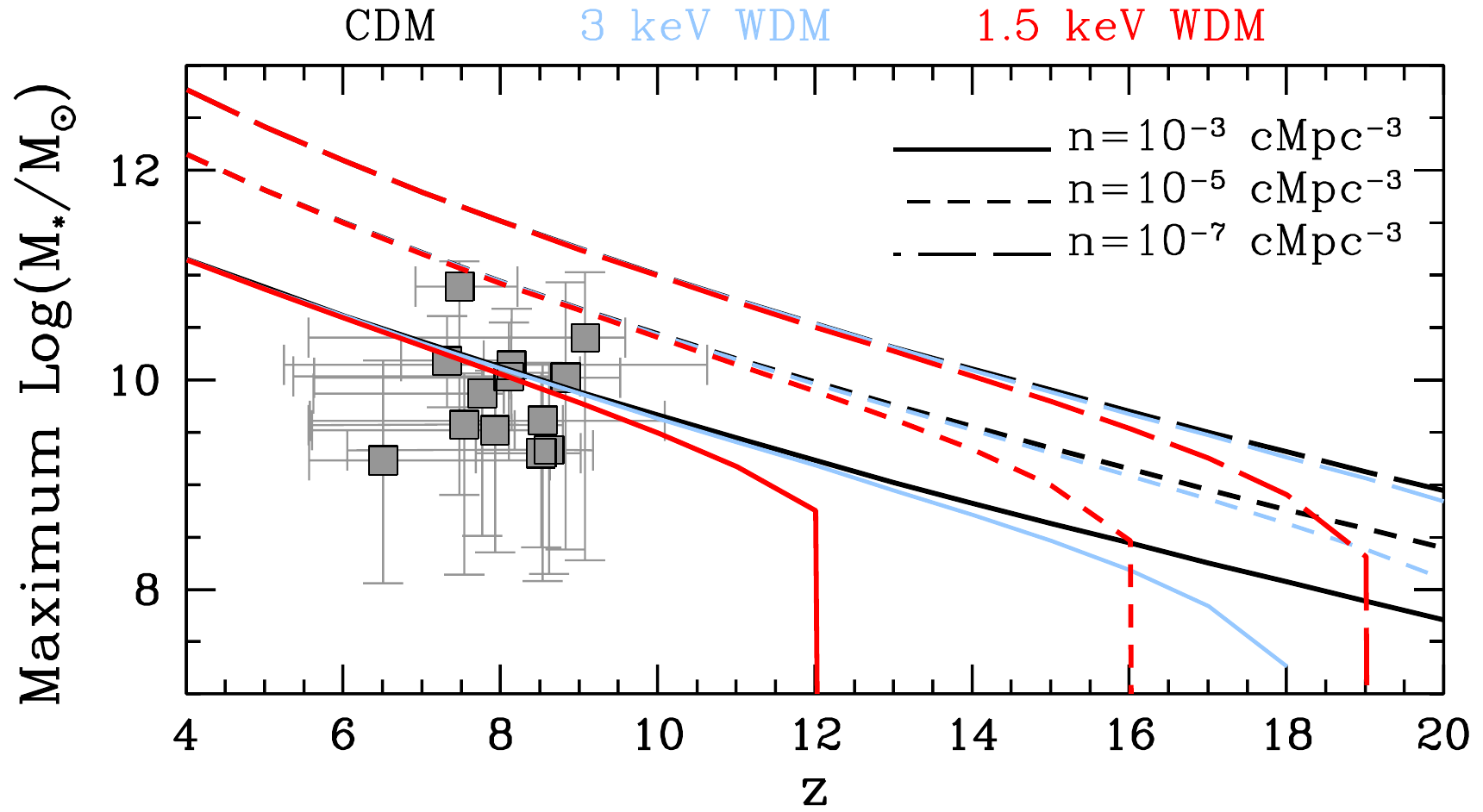}} 
\caption{As a function of redshift, we show the maximum stellar mass allowed for CDM (black lines), 3 KeV WDM (blue lines) and 1.5 keV WDM (red lines). The different lines show results for number densities of $10^{-3} \cmpct$ (solid lines), $10^{-5} \cmpct$ (short-dashed lines) and $10^{-7} \cmpct$ (long-dashed lines), as marked. The data points show the stellar masses derived by \citet{labbe2023} using JWST Cosmic Evolution Early Release Science (CEERS) program data; the error bars account for both the random and systematic uncertainties quoted in their table 2.  }
\label{fig_maxms}
\end{figure*}
\section{The stellar masses of early galaxies - constraints on the WDM particle mass}
We now use the star formation efficiency models above to obtain the evolving SMF at $z \sim 6-15$, the redshift evolution of the associated SMD and the maximum stellar mass expected for a given number density (or volume) cut, as detailed in what follows. 

\subsection{The redshift evolution of the stellar mass function and stellar mass density}
\label{smf}
We show the evolving SMF between $z \sim 6-15$ for CDM, 3 and 1.5 keV WDM in Fig. \ref{fig_smf}. Starting with Model A, where $\epsilon_* = fn(z)$, independent of the DM model used, matching to the massive-end ($M_* \gsim 10^9\msun$ at) of the observed SMF requires a star formation efficiency that decreases with $z$. Indeed, we find $\epsilon_* \sim 15\%$ at $z \sim 6$ which decreases to $\epsilon_* \sim 6\%$ by $z \sim 9$, in accord with theoretical expectations of $\epsilon_* \sim 8-10\%$ at these redshifts \citep[e.g.][]{dayal2014a, dayal2022}. Matching to observations at the low-mass end ($M_* \lsim 10^9 \msun$) in the CDM and 3 keV WDM models requires $\epsilon_*$ to progressively decrease with decreasing mass. Physically this is driven by a combination of the shallow potential wells of such low-mass haloes that limit their star formation efficiency \citep[e.g.][]{dayal2014a} as well as supernova and reionization feedback that could have suppressed their gas masses well below the cosmological ratio \citep[e.g][]{choudhury2019, hutter2021}. Due to a lack of low-mass haloes, the 1.5 keV WDM model shows a turnover at about $10^8\msun$ at $z \sim 6-9$ that shifts to about $10^7 \msun$ by $z \sim 15$. 

In the ``maximal $\epsilon_*$" case (model B), the SMFs from both the 3 and 1.5 keV WDM scenarios converge to CDM for massive galaxies, with $M_* \gsim 10^{10}\msun$. Further, as might be expected, the CDM model over-predicts the observed SMF at all $z \sim 6-15$. As for the WDM models, they show a progressive deficit of lower-mass systems with decreasing $\mx$. For example, the 1.5 (3) keV SMF start peeling away from the CDM SMF at $M_* \sim 10^{9.5} ~(10^{8.5}) \msun$ at all $z \sim 6-15$.

In terms of observations, lensing magnifications have allowed the SMF to be mapped out to masses as low as $10^{6.5-7}\msun$ at $z \sim 6-7$ and  $10^{7}\msun$ at $z \sim 8-9$ \citep{bhatawdekar2019, kikuchihara2020}. As seen from Fig. \ref{fig_smf} even allowing for error bars, the amplitude of the low-mass end of the observed SMF ($\lsim 10^7 \msun$) is only marginally consistent with Model A noted above and is already above the limits predicted by the ``maximal" model at $z\sim 6-7$ in the 1.5 keV WDM scenario. Further, these observational data points approach the upper limits of the 3 keV model, at least at $z \sim 6-7$. While at face value, this might be interpreted as ruling out WDM as light as 1.5 keV, this result must be treated with caution. We note that the observed SMFs collected by the different groups noted \citep[renormalised to a Salpeter IMF by][]{stefanon2021} agree exceedingly well over 2.5 orders of magnitude in mass ($M_* \sim 10^{7.5-10}\msun$) at $z \sim 6-8$; such agreement is not an obvious expectation, given the different data-sets and methodologies used to infer the stellar mass. Indeed, the lensing amplification has been used to reach down to stellar masses as low as $M_* \lsim 10^7 \msun$ \citep{bhatawdekar2019, kikuchihara2020} have a number of associated systematics that might propagate into estimates of the stellar mass \citep{bouwens2017b, atek2018}. Further, all of the above works used parameterised star formation histories that are either constant or increase/decrease as a function of time. However, this assumption can lead to stellar masses being underestimated by as much as a factor of 10, especially for low-mass objects, as has been shown using non-parametric star formation histories to model Atacama Large Millimetre Array (ALMA) REBELS (Reionization Era Bright Emission Line Survey) data at $z \sim 7$ \citep{topping2022}. In any case, 
robust SMFs at $M_* \lsim 10^{7}\msun $ offer an interesting pathway to constraining $\mx$ 
at these early epochs.

We now discuss the associated SMD for both $\epsilon_*$ models, exploring the values for all galaxies and integrating above stellar mass limits of $10^{8}$ and $10^{11}\msun$, as shown in Fig. \ref{fig_smd}. Starting with Model A for the star formation efficiency, integrating over all galaxies CDM predicts an SMD value that drops by about three orders of magnitude, from $10^{8.8}$ to $10^{5.7}\msun \cmpct$ between $z \sim 4$ and 20. Limiting the integration to $M_*\gsim 10^8\msun$ systems results in a much steeper drop of the SMD, from $10^{8.3}$ to $10^{3}\msun \cmpct$ between $z \sim 4$ and 13. The over-prediction of the low-mass end of the SMF with this model results in a similar over-prediction when compared to the observed SMD by about 0.5-0.75 dex. Finally, we find that rare, massive systems with $M_* \gsim 10^{11}\msun$ contain only about 0.6\% of the total SMD by $z \sim 4$. As might be expected, compared to CDM, the SMD values for the 1.5 and 3 keV models show the largest differences (which increase with increasing redshift) when integrating overall systems. In this case, the 1.5 (3) keV models show a fourth (half) of the SMD of CDM at $z \sim 4$ which drops by 3.6 (2) orders of magnitude by $z \sim 14$. We note that the SMDs in CDM and 3 keV WDM effectively show no difference integrating for $M_* \gsim 10^8\msun$ systems; while the 1.5 keV model has about a third less SMD compared to CDM for an integration limit of $M_* \gsim 10^8\msun$, the values all converge when considering the most massive systems with $M_* \gsim 10^{11}\msun$.

This qualitative picture remains unchanged in the ``maximal $\epsilon_*$ model" (panel b of the same figure) which yields an upper limit to the SMD in any of the DM scenarios considered. In this case, integrating above $10^8\msun$, we predict a maximum SMD value that decreases from $10^{9.1}$ to $10^{5.1} \msun \cmpct$ from $z \sim 4$ to 17 for CDM; over the same redshift range, the 1.5 keV WDM model shows a much steeper decline from $10^{8.8}$ to $10^{2} \msun \cmpct$. This offers a first test of the nature of DM using early galaxies: for example, integrating above $10^8\msun$, an observationally inferred SMD value that exceeds $10^7~(10^{7.5}) \msun \cmpct$ at $z \sim 10$ could be used to rule out WDM as light as 1.5 (3) keV.

\subsection{The redshift evolution of the maximal stellar masses in different DM models}
\label{maxms}
We now show the maximum stellar mass allowed (using $\epsilon_*=1$) for a given number density (or volume) for the different DM models, in Fig. \ref{fig_maxms}. Given our assumption of each dark matter halo hosting one galaxy, the number density values quoted correspond to the number density of the host halo. Considering a number density of $10^{-3} \rm {cMpc^{-3}}$, CDM yields a maximum stellar mass of $M_* \sim 10^{11.1} ~ (10^{7.7})\msun$ at $z \sim 4 ~ (20)$. Considering lower number densities of $10^{-7} \rm {cMpc^{-3}}$, these values increase to $M_* \sim 10^{12.8} ~ (10^{8.9})\msun$ at $z \sim 4 ~ (20)$. JWST CEERS observations have been used to assemble a sample of 13 massive galaxies with $M_* \sim 10^{9.2-10.4}\msun$ at $z \sim 6.5-9.08$ \citep{labbe2023} as shown in the same figure. The survey area (40 arcmin$^2$) and redshift uncertainty ($\Delta z\sim 1$) yield a number density of about $10^{-4.9} \rm {cMpc^{-3}}$ for these sources. Within error bars, all of these sources are below the upper limits predicted by CDM for a number density of $10^{-5} \rm {cMpc^{-3}}$, offering a sanity check to our calculations. 

As might be expected, for a given number density, these upper limits to the stellar mass are in accord for all three DM models at $z \lsim 8$. However, the progressive lack of structure formation with increasing redshifts leads to a ``cut-off" of the maximum stellar mass values allowed in light WDM models - for example, with a number density of $10^{-3} \rm {cMpc^{-3}}$, 1.5 KeV WDM models do not predict bound objects at $z \gsim 12$. Decreasing number density limits (i.e. larger volumes) of $10^{-5}$ and $10^{-7}\rm {cMpc^{-3}}$ result in a cut-off at increasing redshifts of $z \sim 16$ and $19$, respectively. These results imply that, for a given number density (volume) limit, the presence of objects above a certain stellar mass could be used to rule out light WDM models. Given the expected impact of cosmic variance in the small JWST fields-of-view \citep[e.g.][]{ucci2021}, one would ideally like to carry out this experiment across multiple blank fields: for example, for a number density of $10^{-3} \rm {cMpc^{-3}}$, the detection of systems, in multiple blank fields, with measurable stellar mass at $z \gsim 12$ (18) could be used to rule out 1.5 (3) keV WDM. 

\section{Conclusions and discussion}
\label{conc}
The nature of DM, and especially limits on the WDM particle mass ($\mx$), remain a key outstanding question in the field of physical cosmology. In this work, we use model-independent phenomenological calculations to revisit WDM particle mass constraints in view of the unprecedented data sets being yielded by the JWST. We use HMFs 
in CDM and WDM ($\mx = 1.5$ and 3 keV) cosmologies to calculate the stellar masses associated with any halo assuming a cosmological baryon-to-DM ratio. We explore two models for the star formation efficiency ($\epsilon_*$): (i) in model A, $\epsilon_*$ is chosen to match to the massive-end of the observed SMF 
at $z \sim 6-10$; (ii) in model B, we use $\epsilon_*=1$ to calculate the ``maximal" stellar masses associated with any halo. Our key findings are:
\begin{itemize}

\item Independent of the DM model considered, matching to the massive-end ($M_* \gsim 10^9\msun$) of the SMF requires $\epsilon_*$ to decrease with $z$, from about 15\% at $z \sim 6$ to 6\% by $z \sim 9$. Matching to the faint end requires an $\epsilon_*$ value that decreases with decreasing mass, hinting at the role of feedback in decreasing the gas masses and/or star formation efficiencies of low-mass haloes.

\item Lensing magnifications (that must be treated with caution) are allowing stellar masses to be probed down to $10^{6.5-7}\msun$ at $z \sim 6-9$. The observed number densities for these objects are already in tension with 1.5 keV WDM models, that lack such low-mass systems. 

\item Integrating over systems more massive than $10^8\msun$, we predict SMD values that show a steep decrease from $10^{8.8}$ to $10^{2} \msun \cmpct$ from $z \sim 4$ to 17 for 1.5 keV WDM; the corresponding values range between $10^{9.1}$ to $10^{5.1} \msun \cmpct$ for CDM. Observed SMD values above the limits predicted by the 1.5 keV WDM model would offer a strong constraint on the minimum $\mx$ value allowed by high-$z$ observations.

\item Finally, we calculate the maximum stellar mass allowed (using $\epsilon_*=1$) for a given number density (or volume) for the different DM models considered. Given the dearth of low-mass haloes, the 1.5 keV WDM model predicts a sharp and earlier cut-off in such maximum masses as compared to the 3 keV or CDM models. For example, with a number density of $10^{-3} \rm {cMpc^{-3}}$, 1.5 (3) KeV WDM models do not predict bound objects at $z \gsim 12$ (18). At this number density, the detection of stellar masses in multiple blank fields $z \gsim 12$ (18) could therefore be used to rule out 1.5 (3) keV WDM. 

\end{itemize}

Over the near future, JWST programs such as CEERS (PI: Finkelstein), Cosmos-Webb (PI: Kartaltepe), PANORAMIC (PIs: Willaims and Oesch) and PRIMER (PI: Dunlop) will be crucial in carrying out our proposed experiment. Although less stringent than the lower-redshift limits of $\mx \gsim 5.2$ keV \citep[e.g.][]{hsueh2020, gilman2020, enzi2021, nadler2021}, such experiments will be crucial in extending constraints on $\mx$ to an era inaccessible by any other means. 

\section*{Data Availability}
Data available on request to the corresponding author.


\section*{acknowledgments}
PD acknowledges support from the NWO grant 016.VIDI.189.162 (``ODIN") and from the European Commission's and University of Groningen's CO-FUND Rosalind Franklin program. The authors thank Atrideb Chatterjee, Paola Santini and Mauro Stefanon for their inputs and data sharing and Ivo Labbe and Laura Pentericci for discussions. Nordita is supported in part by NordForsk.
\bibliographystyle{mnras}
\bibliography{wdm}

\label{lastpage} 
\end{document}